 
 \font\titolino=cmbx10 \font\tsnorm=cmr10 \font\tscors=cmti10
 \font\tsnote=cmr9  
 \font\tscors=cmti10 \font\tscorsp=cmti10
 \magnification=1200
 \hsize=154truemm \hoffset=5truemm \parindent 8truemm \parskip 3 truemm
 plus 1truemm minus 1truemm
 \nopagenumbers
 \newcount\notenumber
  \def\note{\advance\notenumber by 1
 \footnote{$^{\the\notenumber}$}}
 \def\beginref{\bigskip
 \leftline{\titolino References.} \nobreak\noindent}
 \def\ref#1#2{\noindent\item{\hbox to 25truept{[#1]\hfill}}
 #2.\smallskip}
 \def\beginsection#1#2{
 \bigskip \leftline{\titolino
 #1. #2.} \nobreak\medskip\noindent}
 \def\beginappendix#1{
 \bigskip
 \leftline{\titolino Appendix. #1.} \nobreak\medskip\noindent}
 \def\beginack {\bigskip \leftline{\titolino Acknowledgments}
 \nobreak\medskip\noindent}

  \def\sc{\scriptstyle}
 
 \def\Schw{Sch\-warz\-sch\-ild }
 
  \def\da{\dagger}
 \def\pf{\pi_f} \def\pphi{\pi_\varphi} \def\pf{\pi_f} 
 \def\dtvf{\dot\varphi}  
 \def\pz{\pi^0} \def\pu{\pi^1} \def\d{\partial} 
 \def\vf{\varphi}  \def\gmn{g_{\mu\nu}} \def\gmnh{g^{\mu\nu}}
 \def\emn{\eta^{\mu \nu}}  \def\eab{\eta^{\alpha
 \beta}} \def\eabb{\eta_{\alpha \beta}}
 \def\epsab{\varepsilon^{\alpha\beta}} \def\real{{\vrule height 1.6ex
 width 0.05em depth 0ex \kern -0.06em {\rm R}}}
 
 \def\sqr#1#2{{\vcenter{\hrule height.#2pt \hbox{\vrule width.#2pt
 height#1pt\kern#1pt \vrule width.#2pt} \hrule height.#2pt}}}
 \def\bar{\overline}
 \def\AP#1#2#3{{\tscors Ann.\ Phys.} {\bf #1}, #2 (#3)}

 \def\PRD#1#2#3{{\tscors Phys.\ Rev.} {\bf D#1}, #2 (#3)}
 \def\PRL#1#2#3{{\tscors Phys.\ Rev.\ Lett.} \ {\bf #1}, #2 (#3)}
 \def\NPB#1#2#3{{\tscors Nucl.\ Phys.} {\bf B#1}, #2 (#3)}
 \def\NPS#1#2#3{{\tscors Nucl.\ Phys.\ Suppl.} {\bf B#1}, #2 (#3)}
 
 \def\PLB#1#2#3{{\tscors Phys.\ Lett.} {\bf B#1}, #2 (#3)}

 \def\IJMPA#1#2#3{{\tscors Int.\ J.\ Mod.\ Phys.} {\bf A#1}, #2 (#3)}
 \def\IJMPD#1#2#3{{\tscors Int.\ J.\ Mod.\ Phys.} {\bf D#1}, #2 (#3)}
 \def\MPLA#1#2#3{{\tscors Mod.\ Phys.\ Lett.} {\bf A#1}, #2 (#3)}

 \def\ANP#1#2#3{{\tscors Ann.\ Physics (N.Y.)} {\bf #1}, #2 (#3)}
 
 %
 \def\BOL{T.\ Banks and M.\ O'Laughlin, \NPB{362}{649}{1991}}
 \def\KRV{K.V.\ Kucha\v r, J.D.\ Romano and M.\ Varadarajan,
 \PRD{55}{795}{1997} and references therein}
 \def\BH{M.\ Cavagli\`a, V.\ de Alfaro and A.T.\ Filippov, 
 \IJMPD{4}{661}{1995};  \IJMPD{5}{227}{1996}}
 \def\Land{L.D.\ Landau and E.M.\ Lifshitz, {\it
 The Classical Theory of Fields} (Pergamon Press, 1962)}
 \def\Jauch{See
 for instance: J.M.\ Jauch and F.\ Rohrlich, {\it The Theory of Photons
 and Electrons} (Addison-Wesley, 1955) p.\ 103}
 \def\LGK{D.\ Louis-Martinez, J.\ Gegenberg and G.\ Kunstatter, \PLB
 {321}{193}{1994}; D.\ Louis-Martinez and G.\ Kunstatter,
 \PRD{52}{3494}{1995}}
 \def\BJL{E.\ Benedict, R.\ Jackiw and H.-J.\ Lee,
 \PRD{54}{6213}{1996}}
 \def\CJZ{D.\ Cangemi, R.\ Jackiw and B.\ Zwiebach, \ANP{245}{408}{1995}}
 \def\CJA{D.\ Cangemi and R.\ Jackiw, \PRL {69}{233}{1992}}
 \def\CJB{\PRD{50}{3913}{1994}}
 \def\CJC{\PLB{337}{271}{1994}}
 \def\AER{D.\ Amati, S.\ Elitzur and E.\ Rabinovici, \NPB{418}{45}{1994}}
 \def\Kumb{W.\ Kummer and S.R.\ Lau, \AP{258}{37}{1997}}
 \def\Fil{A.T.\ Filippov, \MPLA{11}{1691}{1996}; \IJMPA{12}{13}{1997}}
 \def\CGHS{C.\ Callan, S.\ Giddings,
 J.\ Harvey and A.\ Strominger, \PRD {45}{1005}{1992};
 H. Verlinde, in {\it Sixth Marcel Grossmann Meeting on General
 Relativity}, M. Sato and T. Nakamura, eds. (World Scientific, Singapore,
 1992)}
 \def\Nav{J.\ Cruz and J.\ Navarro-Salas, \MPLA{12}{2345}{1997}; J.\
 Cruz, J.M.\ Izquierdo, D.J.\ Navarro, and J.\ Navarro-Salas, ``{\it Free
 Fields via Canonical Transformations of Matter Coupled
 2d Dilaton Gravity Models}'', Preprint FTUV-97-10, e-Print Archive:
 hep-th/9704168}
 \def\KK{K.V. Kucha\v r, \PRD{50}{3961}{1994}}
 \def\Strom{A.\ Strominger, "{\it Les Houches Lectures on
 Black Holes}", e-print Archive: hep-th/9501071}
 \def\Strobl{T.\ Strobl,
 Proc. of the Second Meeting on Constrained
 Dynamics and Quantum Gravity,  \NPS{57}{330}{1997} } 
 \def\Jacksm{R.\ Jackiw,
 Proc. of the Second Meeting on Constrained
 Dynamics and Quantum Gravity,  \NPS{57}{160}{1997} } 
 %
 \null 
 \vskip 20truemm
 \centerline{\titolino QUANTIZATION OF THE STRING INSPIRED DILATON 
  GRAVITY}
 \vskip 3truemm
 \centerline{\titolino AND THE BIRKHOFF THEOREM}
 \vskip 9truemm
 \centerline{\tsnorm Marco Cavagli\`a$^{(a)}$,
  Vittorio de Alfaro$^{(b, d)}$ and Alexandre T. Filippov$^{(c)}$}
 \bigskip
 \centerline{$^{(a)}$\tscorsp Max-Planck-Institut f\"ur
  Gravitationsphysik,}
 \smallskip
 \centerline{\tscorsp Albert-Einstein-Institut,}
 \smallskip
 \centerline{\tscorsp Schlaatzweg 1, D-14473 Potsdam, Germany.}
 \bigskip
 \centerline{$^{(b)}$\tscorsp Dipartimento di Fisica
  Teorica dell'Universit\`a di Torino, }
 \smallskip
 \centerline{\tscorsp Via Giuria 1, I-10125 Torino, Italy.}
 \bigskip
 \centerline{$^{(c)}$\tscorsp Joint Institute for Nuclear Research}
 \smallskip
 \centerline{\tscorsp R-141980 Dubna, Moscow Region, Russia.}
 \bigskip
 \centerline{$^{(d)}$\tscorsp INFN, Sezione di Torino, Italy.}
 \vskip 8truemm
 \centerline{\tsnorm ABSTRACT} \begingroup\tsnorm\noindent
 We develop a simple scheme of quantization for the dilaton CGHS model
 without scalar fields, that uses the  Gupta-Bleuler approach for the
 string fields.  This is possible because the constraints can be
 linearized classically, due to positivity conditions that are present in
 the model (and not in the general string case). There is no ambiguity
 nor anomalies in the quantization. The expectation values of the metric
 and dilaton fields obey the classical requirements, thus  exhibiting
 at the quantum level the Birkhoff theorem.
 %
\smallskip
\noindent
{\tsnorm PAC(S): 04.60.-m, 04.60.Kz, 03.70.+k.\hfill}
\break\noindent
{\tsnorm Keyword(s):} {\it Quantum Gravity, Two-Dimensional models,
 Canonical Quantization, Field Theory}
\vfill
\hrule
\noindent
\leftline{E-Mail: cavaglia@aei-potsdam.mpg.de\hfill}
\leftline{E-Mail: vda@to.infn.it\hfill}
\leftline{E-Mail: filippov@thsun1.jinr.dubna.su\hfill}
\endgroup
\vfill
\eject
\footline{\hfill\folio\hfill}
\pageno=1
\noindent
%
\beginsection{1}{Introduction}
Recently, much work has been done on gravitational models in 1+1
dimensions, the solution of which may hopefully clarify some difficult
problems of quantum black holes and, more generally, of quantum gravity in
higher dimensions [1].  Most of the models studied in some detail in the
literature are special cases of the general dilaton gravity coupled to
gauge and scalar fields (for a very compact overview and references see
[2]). In general, these models are not integrable and their solution
cannot be found even classically. However, interesting models which are
classically integrable are also known whose solutions can be explicitly
written in terms of free massless fields (the most general class of such
models is presented in [2]). Unfortunately, not much is known about their
quantization: only the so-called CGHS model [3] has been studied in detail
(see Refs.\ [4,5,6]; for a review of earlier results see also [1]).  These
investigations revealed two main obstacles to quantization: first, the
quantum canonical transformations to free fields are highly nontrivial and
difficult to construct; second, the quantization is obstructed by
anomalies in the commutator of the constraints, and the theory needs to be
modified. In the CGHS model, the canonical transformations and
modification of the constraints have recently been implemented [5,6]. 
However, the quantization of more general integrable models is a
completely open problem.

The problems mentioned persist even in the dilaton gravity models not
coupled to scalar matter fields. The general dilaton gravity model is
[7,8,2]
$${\cal L} = \sqrt{-g} [U(\vf) R(g) + V(\vf) +
W(\vf) \gmnh \d_{\mu} \vf \d_{\nu} \vf]\,,\eqno(1)$$
where $\mu,\nu=0,1$; $U$, $V$, $W$ may be arbitrary functions of
$\varphi$, and $R$ is the scalar curvature. Locally, we may always choose
$U = \vf$ (or, $U=\exp(c \vf$), etc.).  Using the Weyl rescaling, $\gmn =
\Omega(\vf) {\bar{g}}_{\mu \nu}$, we may make $W$ equal to any given
function (e.g.\ $\bar W=0$). The CGHS model without matter is equivalent,
in this sense, to the theory with $W = 0$ and $V = {\rm const}$ (with $V =
\vf$ we get the Jackiw-Teitelboim model). 

The general dilaton gravity described by Eq.\ (1)  is integrable and both
the metric and dilaton fields may be expressed in terms of one free
(D'Alembert) field and of one invariant parameter which is a local
integral of motion independent of the coordinates. (For the \Schw black
hole, it is the black hole mass.) Using the gauge in which this free field
is one of the coordinates, one finds that all solutions depend on one
coordinate. This means that the metric can be explicitly expressed in
terms of the dilaton and thus has at least one horizon. (For more details
see, e.g.\ Ref.\ [2].) These properties constitute a generalization of the
classical Birkhoff theorem for spherically symmetric gravity in any
dimension (we thus call, somewhat improperly, ``static'' these solutions).

The drastic reduction of the dilaton gravity field theory to a
finite-dimensional dynamical system signals that the general dilaton
gravity is actually a topological theory.\note{\tsnote The topological
nature of the CGHS and Jackiw-Teitelboim models is well-known: they are
topological BF theories (see [4,9,5]). For the topological formulation of
the general dilaton gravity see [10] and references therein.} The Birkhoff
theorem is also valid for the dilaton gravity (1) coupled to Abelian gauge
fields ([8,2]). However, any coupling to scalar fields destroys the
topological nature of the theory and invalidates the Birkhoff theorem.
Moreover, 0+1 dimensional solutions of dilaton gravity coupled to scalar
matter have no horizons (the ``no horizon theorem'' [2]).  As a
consequence, the coupling to scalar fields cannot be treated
perturbatively, even in the classical theory. 

Unlike the theories coupled to scalars, the general dilaton gravity model
can be quantized by first reducing it to a dynamical system with a single
constraint. (This may be regarded as a particular gauge fixing.) The
quantization of the finite-dimensional system so obtained is more or less
straightforward, and the resulting Hilbert space is spanned by the
eigenvectors of the (gauge invariant) mass operator. This quantization
approach, that uses first the Birkhoff theorem and then the quantization
algorithm, has been introduced in Ref.\ [11] in the case of the 3+1
dimensional Schwarzschild black hole; the procedure can be generalized to
any pure two dimensional dilaton gravity model.\note{\tsnote It is not
difficult to rewrite all the formulas of Ref.\ [11] for the simpler CGHS
case. The treatment of more general potentials $\sc V(\varphi)$, when many
horizons may exist, requires a more careful consideration.}

A further approach to the quantization of the pure dilaton gravity models
does not use the topological nature of the models nor the Birkhoff
theorem. Since the classical solution can be written as a function of a
single free field, one may try to find a canonical transformation of
dilaton gravity to constrained free fields. This transformation is known
for some time for the CGHS case [5,6], and recently an interesting new
proposal in this direction has been advanced [12]. However, up to now an
explicit canonical transformation has not been constructed for the general
case and this approach was worked out only for the CGHS model which was
the subject of deep investigations (with and without scalar matter, see
Refs.\ [4,5,6]). 

The main results of these investigations are clearly summarized in the
report [13]. Although a beautiful canonical transformation which
linearizes the CGHS model and represents it in terms of an infinite
bosonic string does exist, the quantization is not straightforward even in
the absence of scalar fields. Indeed, due to the presence of anomalies,
different quantum theories corresponding to the same classical model can
be found. In particular, in the Schr\" odinger representation there exists
a quantization scheme in which the anomaly in the pure dilaton gravity is
cancelled [5]. However, this scheme is based on the use of negative energy
states for the string fields instead of the standard string quantization
that introduces negative norm states, and the result obtained has no
evident connection to the quantum version of the reduced theory via the
Birkhoff theorem.

In this paper we consider a further scheme for the pure dilaton CGHS
model, that directly proves the Birkhoff reduction at the quantum level
and does not produce any anomalies. The quantization is carried out, by
use of the standard Gupta-Bleuler method, on the string variables. This is
possible because the model has actually more information in it than the
canonically transformed version described by a couple of string fields. In
fact, we will argue that this model is not equivalent to a string theory.

In the first place, there exists a gauge invariant local mass that cannot
be expressed purely in terms of the string fields. Moreover, further
information comes from the canonical transformation between the original
fields and the string ones, under the form of positivity conditions for
certain functions. This allows an essential step: the linearization of the
constraints. Let us stress that the linearized constraints generate
reparametrizations of the metric. There is no need to consider the
original quadratic constraints as fundamental operators.

Hence, it becomes possible to quantize the model with the standard choice
of the quantum vacuum by the Gupta-Bleuler method. There is no anomaly in
the algebra of the linearized constraints (the algebra of the quadratic
constraints does not concern us any more). This quantization of the
string-like fields is essentially equivalent to the ``Schr\"  odinger''
quantization that uses positive norms and negative energy states [5]. 

The only gauge invariant operators are the mass and its conjugate momentum
(conjugate operators were discussed in Refs.\ [14,8]). The ground state of
the string must be labeled by the eigenvalue of the mass operator (as
discussed in a different context in Refs.\ [4,8]).

Our quantization procedure explicitly shows how the reduction to quantum
mechanics is achieved by the Gupta-Bleuler quantization of the field
theory. With linearized constraints, the quantized field theory has the
same content as scalar-longitudinal electrodynamics and is pure gauge. The
set of Gupta-Bleuler states corresponds to reparametrization of the
coordinates, expectation values of the metric and dilaton fields agree
with the corresponding classical quantities and the ``Birkhoff reduced''
quantum theory is recovered.
\beginsection{2}{Action and Hamiltonian Formalism} 
Our starting point is the two-dimensional action related by a Weyl
transformation to the pure dilaton CGHS [3] model
$$S=\int d^2x\, \sqrt{-g}\, \left[\vf R -
{\lambda \over 2} \right]\,, \eqno(2)$$ 
here $\gmn$ is a two-dimensional metric and $\vf$ is the dilaton field
(for $R$ we follow the conventions of [15]). As in [5] we write the
two-dimensional metric as
$$\gmn=\rho\left(\matrix{\alpha^2-\beta^2&\beta\cr
\beta&-1\cr}\right)\, .\eqno(3)$$ 
Here $\alpha(x_0,x_1)$ and $\beta(x_0, x_1)$ play the role of the lapse
function and of the shift vector respectively; $\rho(x_0,x_1)$ represents
the dynamical gravitational degree of freedom. The coordinates $x_0,\,
x_1$ are both defined on $\real$. It is convenient to introduce the
variable $f=\ln \rho$. Using (3) the action (2) can be written in the
Hamiltonian form (see e.g.\ [5]) as\note{\tsnote Note that all the
formulae of this section can be easily rewritten for the general dilaton
gravity (1).}
$$S=\int d^2x\,
\left[\dot f\pi_{f}+\dtvf\pphi-\alpha {\cal H}-\beta{\cal P}\right]\, ,
\eqno(4)$$ 
where $\pi_f$ and $\pphi$ are the conjugate momenta to $f$
and $\vf$ respectively, and $\cal H$ and $\cal P$ are the constrained
super-Hamiltonian and super-momentum: 
$$\eqalignno{&{\cal
H}=\pf\pphi+f'\vf'-2\vf''+{\lambda\over 2} e^f\, =\,0\, , &(5a)\cr &{\cal
P}=2\pf'-\pphi\vf'-\pf f'\, =\,0\, .&(5b)\cr}$$ 
We may also define a functional $M$ of the canonical variables which is
conserved under time and space translations (analogous to the \Schw\ mass)
[8,2,16]. In our notations $M$ is given by
$$M~=~{\lambda\over 2} \vf ~+~e^{-f}(\pf^2-\vf'^2)\,.\eqno(6)$$
It is straightforward to prove that $\dot M=M'=0$ using the equations of
motion and the constraints. 

The Birkhoff reduction to static configuration (depending on one
coordinate) may be stated as follows. We may set $\alpha=1$ and $\beta= 0$
and introduce the coordinates
$$u={1\over 2}(x_0+x_1)\,,
~~~v={1\over 2}(x_0-x_1)\,;\eqno(7)$$
the two-dimensional line element corresponding to the metric tensor (3)
becomes
$$ds^2=4\rho(u,v)dudv\,.\eqno(8)$$
A metric of this form is static if and only if $\rho$ can be cast in
the form [2]
$$\rho(u,v)=h(\Psi){da(u)\over du}\, {db(v)\over dv}, ~~~~~
\Psi\equiv a(u)+b(v)\,,\eqno(9)$$
where $a$ and $b$ are arbitrary functions. This metric depends on one
coordinate. If, in addition, $\vf$ depends only on $\Psi$, the solution
will be called static. To see that all the solutions of (2) are static we
write the constraints and the equations of motion in the coordinates (7) 
[2]
$$\eqalignno{
&\d_u\left(e^{-f}\d_u\vf \right)=\d_v\left(e^{-f}\d_v\vf
\right)=0\,,&(10a)\cr
&\d_u\d_v f=0\,,~~~~
\d_u\d_v\vf + {\lambda\over 2} e^f =0\,,&(10b)}$$
where the constraints (10$a$) are valid for the generic model (1)  (but
will be destroyed by adding any coupling to additional scalar fields).
The constraints (10$a$) can be solved in terms of a free field: 
$$\eqalignno{&\rho\equiv e^f={dF\over d\Psi}\d_u\Psi\d_v\Psi\,,&(11a)\cr
&\vf=F(\Psi) \,,&(11b)\cr
&\d_u \d_v \Psi\,=\,0\,.&(11c)}$$
This shows that all the solutions of the generic dilaton gravity (1)  are
static, which is the generalized Birkhoff theorem. The function $F(\Psi)$
is determined by the equations of motion and depends on the potential
$V(\vf)$. For the CGHS model we find $F(\Psi)$ from (10$b$)
$$F(\Psi)=C_0 e^{-\lambda\Psi/2}\, +\, {2M\over \lambda}\,.\eqno(11d)$$
Here $C_0$ is an integration constant; the second integration constant was
expressed in terms of the constant $M$ by using (6). One sees that $M$
appears as a zero mode of the field $\vf$. Now let us recall the canonical
free field formalism that will be the starting point for the quantum
theory. 
\beginsection{3}{String variables and linearization of the constraints}
Let us use the transformation [5]
$$\eqalign{&A_0={2\over\lambda}e^{-f/2}\left(\pf\cosh{\Sigma}-
\vf'\sinh{\Sigma}\right)\,,~~
\pz=-\lambda e^{f/2}\cosh{\Sigma}-\lambda A_1'\,,\cr
&A_1={2\over\lambda}e^{-f/2}\left(\pf\sinh{\Sigma}-
\vf'\cosh{\Sigma}\right)\,,~~
\pu=\lambda e^{f/2}\sinh{\Sigma}+\lambda A_0'\,,\cr}\eqno(12)$$
where
$$\Sigma(x_1)={1\over 2}\int^{x_1}_{-\infty}
dx_1'\pphi(x_1')\,.\eqno(13)$$
The above transformation is canonical for the field variables $(-\vf',
\Sigma, f, \pf; A_{\alpha}, \pi_{\alpha})$. Note that the inverse
transformation only defines $\vf$ up to a zero mode $\vf_0(x_0) \equiv
\vf(x_0,c)$. To make the transformation (12) invertible, one has to
supplement the new field variables $A_{\alpha}, \pi_{\alpha}$ by a pair of
conjugate variables, e.g.\ $\vf(x_0, c=\infty)$ and $2\Sigma(\infty)$
($\vf_0$ will commute with all the field variables if we choose $c =
\infty$).

Using (12) the two constraints become
$$\eqalignno{&{\cal H}={1\over
2\lambda}\pi^\alpha\pi_\alpha+{\lambda\over 2}A'^\alpha A'_\alpha=0\,,
&(14a)\cr
&{\cal P}=-\pi^\alpha A'_\alpha=0\,.&(14b)\cr}$$
Let us introduce the metric and the Levi - Civita tensors as
$$\eta^{\alpha \beta}=\left(\matrix{1&0\cr 0&-1\cr}\right)\,;~~~~~
\epsab=\left(\matrix{0&1\cr -1&0\cr}\right)\,.\eqno(15)$$
The functional $M$ defined in (6) is represented as
$$M=M_0+{\lambda\over 4}\int_{b}^{x_1} dx_1' ~\bigl(\epsab\pi_\alpha
A_\beta~
+~\lambda A^\alpha A'_\alpha\bigr)\,.\eqno(16)$$
(The value of $b$ is irrelevant.) Here $M_0$ must be a constant since $M$
is independent of $t$. The zero mode $\vf_0$ and the zero modes of the
string fields can be related to $M_0$ but we do not need their precise
relationship (see [6]).  On the equations of motion we have $M=M_0$. 
Let us introduce the operators ${\cal C}^\alpha$ and ${\cal D}^\alpha$
$$\eqalign{&{\cal C}^\alpha=
\pi^\alpha-\lambda \varepsilon^{\alpha\beta}A_\beta'\,,\cr
&{\cal D}^\alpha=
\pi^\alpha+\lambda \varepsilon^{\alpha\beta}A_\beta'\,.\cr}\eqno(17)$$
$\dot M$ and $M'$ are proportional to ${\cal
C^{\alpha}}$.
The classical quantities $\cal H$, $\cal P$, $M$, $\rho$, $\vf$ can be
written as functions of ${\cal C}^\alpha$, ${\cal D}^\alpha$ and
$A_{\alpha}$. They read
$$
\eqalignno{&{\cal H}={1\over 2\lambda}{\cal C}^\alpha {\cal D}_\alpha\,,
&(18a)\cr
&{\cal P}={1\over 2\lambda}\varepsilon_{\alpha\beta}{\cal D}^\alpha {\cal
C}^\beta\,, &(18b)\cr
& M = M_0 -{\lambda\over 4}\int^{x_1}_b dx_1'\,\varepsilon^{\alpha \beta}
A_{\alpha}
{\cal C}_{\beta}\,,
&(18c)\cr
&\rho={1\over\lambda^2}{\cal D}^\alpha {\cal D}_\alpha\,, &(18d)\cr
&\varphi={2M \over \lambda} - {\lambda \over 2}
A_{\alpha} A^{\alpha}\,.&(18e) }
$$
The Poisson brackets of ${\cal C}^\alpha$ and ${\cal D}^\alpha$ are
$$[{\cal C}^\alpha(x_0,x),{\cal
D}^\beta(x_0,y)]=2\lambda
\varepsilon^{\alpha\beta}\d_y\delta(x-y)\,.\eqno(19)$$

Now we show that the constraints can be linearized, using a suitable
redefinition of the Lagrange multipliers. The linearized constraints are
the functions ${\cal C}^{\alpha}$.  Thus in this form of the theory the
generators are linear in the string variables $A_{\alpha}$.

It is useful to introduce the conjugate variables [6]
$({\cal P}_{\pm},{\cal X}_{\pm})$ where
$$\eqalign{
&{\cal P}_{-}={1\over {2 \sqrt{\lambda}}}({\cal C}^0-{\cal C}^1)\,,~~~~
{\cal X}_{-}'=-{1\over {2\sqrt{\lambda}}}({\cal D}^0+{\cal D}^1)\,,\cr
&{\cal P}_{+}={1\over 2{\sqrt{\lambda}}}({\cal C}^0+{\cal C}^1)\,,~~~~
{\cal X}_{+}'={1\over {2\sqrt{\lambda}}}({\cal D}^0-{\cal D}^1)\,,\cr}
\eqno(20)$$
and also define the constraints ${\cal H}_+$ and ${\cal H}_-$ that
generate reparametrization in $u$ and $v$; they can be written as
functions of the constraints ${\cal P}_{\pm}$ and of the space derivative
of the conjugate constraints, ${\cal X}'_{\pm}$: 
$$
\eqalign{&{\cal H}_{+}\equiv{\cal H+P}={1\over {2\lambda}}({\cal
D}^0+{\cal D}^1)({\cal C}^0-{\cal C}^1)=-2{\cal X}_-'{\cal P}_-\,,\cr
&{\cal H}_{-}\equiv{\cal
H-P}={1\over {2\lambda}}({\cal D}^0-{\cal D}^1)({\cal C}^0+{\cal
C}^1)=2{\cal X}_+'{\cal P}_+\,.\cr}\eqno(21)
$$
Using these variables the Lagrangian density reads
$${\cal L}=\dot {\cal X}_{+}{\cal P}_{+}+\dot {\cal X}_{-}{\cal P}_{-}-
l_{+}{\cal H}_{+}-l_{-}{\cal H}_{-}\,,\eqno(22)$$
where $l_\pm$ are the suitable combinations of the Lagrange multipliers.
In Appendix we show that ${\cal H}=0$ and ${\cal P}=0$ are satisfied iff
${\cal C}^\alpha=0$ since ${\cal X'}_{\pm}$ have definite signs.
Accordingly, the coefficients ${\cal X}_{\pm}$ can be reabsorbed in the
Lagrange multipliers. 
We redefine the multipliers as
$$
r_+= 2 {\cal X}_+' ~l_-, ~~~~r_- = -2 {\cal X}_-' ~l_+\,.\eqno(25)
$$
The Lagrangian density is now
$$
{\cal L}=\dot {\cal X}_{+}{\cal P}_{+}+\dot {\cal X}_{-}{\cal P}_{-}-
r_{+}{\cal P}_{+}-r_{-}{\cal P}_{-}\,.\eqno(26)
$$

The linear constraints, ${\cal P}_{\pm}$ or ${\cal C}_{\alpha}$, generate
the reparametrizations of the metric. Indeed we have the Poisson brackets
$$
[\rho(x_0,x),{\cal
C}^\alpha(x_0,y)]=4\varepsilon^{\alpha\beta}\lambda^{-1}{\cal
D}_\beta(x_0,x)\d_y\delta(x-y)\,.\eqno(27)
$$

As a consequence of the linearization, the theory can be quantized in a
simple scheme. Before doing that, let us see the form that the classical
solution (11) takes in terms of the fields $A_{\alpha}$. The equations of
motion are
$$\lambda \dot A_\alpha=\eabb\pi^\beta\,,~~~~
\dot\pi^\alpha=\lambda\eab A_\beta''\,\eqno(28)$$
and the solution is
$$ A_\alpha=U_\alpha(u)+V_\alpha(v)\,.\eqno(29)$$
The linear constraints ${\cal C}^{\alpha}=0$  correspond to 
$$\eab \d_{\alpha} A_{\beta}=0,~~\epsab \d_{\alpha} A_{\beta}=0,
~~{\rm or,}~~
U_0(u)=U_1(u),~~V_0(v)=-V_1(v).\eqno(30)$$
It is easy to prove that the solution (29,30) coincides with (11).
Substituting Eqs.\ (29,30) into ($18d,e$) one obtains
$$\eqalignno{&\rho\equiv e^f\, = \, 
4{dU_0(u)\over du}\, {dV_0(v)\over dv}\,, &(31a)\cr
&\vf\,=\,{2M \over \lambda}- 2\lambda U_0(u)V_0(v)\,.&(31b)\cr}$$
This solution coincides with (11) if $\Psi$ is defined by
$$C_0 e^{-\lambda\Psi/2}=-2\lambda U_0(u) V_0(v).\eqno(32)$$
The ``staticity'' of the classical solution is thus embodied in Eqs.\
(31).
\beginsection{4}{Quantization}
The quantization starts from the introduction of the
Lagrangian\note{\tsnote From now on we set $\sc\lambda=1$. There is no
real loss of generality while the formulae become more elegant.}
$${\cal L}= {1 \over 2} \d_{\mu}A_{\alpha} \d_{\nu}A_{\beta}
\emn \eab\,.\eqno(33)$$
The commutation relations are
$$\bigl[ A_{\alpha}(x),A_{\beta}(y) \bigr]= -\eabb
\int {d^2k \over 2\pi} \delta(k^2) \varepsilon(k_0)
e^{ik(x-y)}\,. \eqno(34)$$
The field expansion is
$$A_{\alpha}= \int_{-\infty}^\infty {dk\over 2\sqrt{\pi \omega}}
\bigl\{b_{\alpha}(k)e^{-i\omega x_0+ikx_1}
+b^\dagger_{\alpha}(k)e^{i\omega x_0- ikx_1}\bigr\}\,,\eqno(35)$$
where $\omega=|k|$. From (29) we obtain
$$\eqalign{&U_{\alpha}= \int_0^\infty {dk \over 2\sqrt{\pi k}}
\bigl\{ a_{\alpha}(k)e^{-2iku} +a^\dagger_{\alpha}(k)e^{2iku}\bigr\}\,,\cr
&V_{\alpha}=\int_0^\infty {dk \over 2\sqrt{\pi k}}
\bigl\{b_{\alpha}(k)e^{-2ikv}
+b^\dagger_{\alpha}(k)e^{2ikv}\bigr\}\,,\cr}\eqno(36)$$
and $a_\alpha(k)=b_\alpha(-k)$, $k>0$. Consequently ($k>0$) the
non-vanishing commutators are
$$\bigl[a_{\alpha}(k),a^\dagger_{\beta}(k')\bigr] = \eabb
\delta(k-k')\,,~~~~
\bigl[b_{\alpha}(k),b^\dagger_{\beta}(k')\bigr] = \eabb
\delta(k-k')\,.\eqno(37)$$
It follows that
$$\bigl[U_{\alpha}(u_1),U_{\beta}(u_2)\bigr] =
-{i\over 4} \eabb \varepsilon(u_1-u_2),~~~
\bigl[V_{\alpha}(v_1),V_{\beta}(v_2)\bigr] =
-{i\over 4} \eabb \varepsilon(v_1-v_2).\eqno(38)$$
The canonical quantities $({\cal P}_\pm, {\cal X}_\pm)$ can be expressed
in function of $q_{a, b}(k)$ and $p_{a, b}(k)$, where
$$\eqalignno{
&q_a=a_0-a_1\,,~~~~~~q_b=b_0+b_1\,,&(39a)\cr
&p_a=a_0+a_1\,,~~~~~~p_b=b_0-b_1\,.&(39b)}
$$
Their non vanishing commutators are
$$
\bigl[q_a(k), p_a^{\dagger}(k')\bigr] =  \bigl[q_b(k),p_b^{\dagger}(k')\bigr]=
2 \delta(k-k')\,. \eqno(40)
$$
We have
$$
\eqalign{&{\cal P}_{+}=-i\int_0^\infty {dk k\over 2\sqrt{\pi
k}}\left(q_a(k)e^{-2iku}-q_a^\dagger(k)e^{2iku}\right)\, , \cr
&{\cal P}_{-}=-i\int_0^\infty {dk k\over 2\sqrt{\pi
k}}\left(q_b(k)e^{-2ikv}-q_b^\dagger(k)e^{2ikv}\right)\, , \cr
&{\cal X}_{+}=\int_0^\infty {dk\over 2\sqrt{\pi
k}}\left(p_a(k)e^{-2iku}+p_a^\dagger(k)e^{2iku}\right)\, , \cr
&{\cal X}_{-}=\int_0^\infty {dk\over 2\sqrt{\pi
k}}\left(p_b(k)e^{-2ikv}+p_b^\dagger(k)e^{2ikv}\right)\, .
\cr}\eqno(41)
$$
The classical quadratic constraints (14,21) cannot be implemented
operatorially, since as operator equations they are in contrast with the
quantization rules (34,37,38) and further they exhibit the usual bosonic
string anomaly, $c=2$ (see e.g.\ [5]). However this does not concern us: 
in the present scheme the generators of reparametrizations are the ${\cal
C}^{\alpha}$ and there is no anomaly for them since they commute. So we
may carry out a different quantization scheme, quantizing the theory with
the linear constraints (17). Then the Gupta-Bleuler procedure can be
carried out following the lines of QED [17].  The vacuum is chosen as
(string vacuum) 
$$a_{\alpha}|0>=0\,,~~~~~~~~b_{\alpha}|0>=0\,.\eqno(42)$$
This leads to negative norm states. Now we implement the constraints by
requiring that, for each oscillation mode, physical states be selected
by
$$q_a |\Psi>\,=\,0\,,~~~~~~ q_b |\Psi>\,=\,0\,.\eqno(43)$$
The states $|\{n_a,n_b\}>$ defined as
$$|\{n_a,n_b\}>\equiv q^{\da}_a(k_1)...q^\da_a(k_{n_a})\,
q^{\da}_b(k'_1)...q^\da_b(k'_{n_b})|0>\eqno(44)$$
satisfy the condition (42) (remember that
$\bigl[q_{a,b}(k),q_{a,b}^\dagger(k')\bigr] =0$), and have zero norm if
$n_a\,\not=0$ or $n_b\,\not=0$.  The general solution of the constraints
(42) is then
$$|\Psi>\,=\,\sum_{n_a,\,n_b}\int d^{n_a}k \int d^{n_b}k'~
C_{n_a n_b}(k_1,...k_{n_a};k'_1,...k'_{n_b}) |\{n_a,n_b\}>\,.
\eqno(45)$$
The norm of this state is
$$<\Psi|\Psi>\,=\,|C_{00}|^2\,. \eqno(46)$$
The  constraints ${\cal H} =0,~{\cal P} = 0$ hold for matrix elements:
$$<\Psi_2|:{\cal H}:|\Psi_1>=0\,,~~~~
<\Psi_2|:{\cal P}:|\Psi_1>=0\,,\eqno(47)$$
where the normal ordering with annihilation operators on the right must be
used (the algebra of ${\cal H}$, ${\cal P}$ shows an anomaly, but this is
irrelevant here, as now our generators of gauge transformations
(reparametrizations) are the constraints ${\cal C }_{\alpha}$ that have
none). 

Using (22) and (44) the expectation value of $\rho$ is
$$<\Psi|:\rho(u,v):|\Psi>= 4 {dF(u)\over du}\, {dG(v)\over
dv}\,,\eqno(48)$$
where
$$\eqalign{&F(u)=\int {dk\over 2\sqrt{\pi k}}\,\left(
C_{00}{}^* C_{10}(k) e^{-2iku}~+~
C_{00} C_{10}(k){}^* e^{2iku}\right)\,,\cr
&G(v)=\int {dk\over 2\sqrt{\pi k}}\,
\left(C_{00}{}^* C_{01}(k) e^{-2ikv}~+~
C_{00} C_{01}(k){}^* e^{2ikv}\right)\,.\cr}\eqno(49)$$
The result (48) is analogous to the classical relation (31$a$); we have
of course
$$F(u)G(v)=<\Psi|U_0(u)V_0(v)|\Psi>\,.\eqno(50)$$
Note that $<\Psi|:\rho(u,v):|\Psi>$ has the form
$$<\Psi|:\rho(u,v):|\Psi>=
h\bigl(a(u)+b(v)\bigr)\,{da(u)\over du}\,{db(v)\over dv}\,,\eqno(51)$$
which is the essence of classical staticity.  Let us now consider the
operator $M$, Eq.\ (16). The quantity $I$ that is the integrand in (16) 
classically vanishes. In the quantum case, each term in $I$ contains one
of the operators $q_{a,b}$ or $q^{\dagger}_{a,b}$. Adopting a normal
ordering, the matrix elements of $I$ between physical states vanish. This
corresponds to the classical property. So,
$$<\Psi_2|M|\Psi_1>=<\Psi_2|M_0|\Psi_1>\,.\eqno(52)$$
Further, the operator $M_0$ commutes with all the creation and annihilation
operators of $A_\alpha$, since $M_0$ is the zero mode of the field
$\vf$. So we must characterize the vacuum by a further quantum number:
$$M_0|0;m>=m|0;m>. \eqno(53)$$
Equation (53) is of high interest. There are infinite vacua, differing by
the eigenvalue of $M_0$. The only gauge invariant label of a state is $m$. 
This result is similar to the case of the \Schw\ metric discussed in [11],
where staticity was imposed from the beginning, reducing the problem to
quantum mechanics, and states were labeled by the eigenvalues of the mass
operator.  Finally, the expectation value of $\vf$ reads
$$\eqalign{<\Psi;m|:\vf(u,v):|\Psi;m>\,&=\,2m-2
<\Psi;m|U_0(u) V_0(v)|\Psi;m>\cr
&=\,2m-2F(u)G(v)}\eqno(54)$$
in analogy to (31$b$).

We conclude with two remarks. The first is that the roles of $A_0$ and
$A_1$ can be interchanged, i.e.\ the sign in Eq.\ (34) can be changed,
because the condition (14$a$) shows that the choice of the ``right''
metric field is irrelevant. The operators $q_a$ and $q_b$ will again
contain one operator with wrong metric and one with right metric;  nothing
changes in the construction (45) of the physical states.  Second remark:
our quantization rule (34) amounts to assuming that $x_0$ is time, namely
that the canonical equal $x_0$ commutators for $A_0$ hold.  The
construction of the physical states is actually independent of which
variable, $x_0$ or $x_1$, is chosen as time in defining the canonical
commutators.  Indeed, let us proceed by canonical equal $x_1$ quantization
for $A_0$.  In that case the rule (37) is suitably modified,  the
commutators of $b_{\alpha}$ change sign: now $b_0$ has wrong metric while
$b_1$ has the correct one.  Again, in $q_b$ there appears one operator
with the right and one with the wrong metric and the construction of
physical states, Eq.\ (45), remains unchanged. 
\beginsection{5}{Conclusions}
We have proposed a new quantization of the model, based on the
linearization of the constraints, that may be performed because in the
present case we possess further information on the theory, due to the
existence of the transformation (12) from the original fields to the
string ones (there is no analogous property for the string). The
linearized constraints generate reparametrizations. It follows then that
the Gupta-Bleuler procedure can be applied in the present case and there
are no anomalies.  Classically, taking into account the constraints (21),
all the field theory tells us is just that there is a single free field
whose degrees of freedom correspond to reparametrization of the
coordinate;  indeed, a choice for $U_0(u),\, V_0(v)$ defines $\Psi$, and
the different choices correspond to different solutions (11). In the
quantum theory, the physics contained in $A_\alpha$ is pure gauge,
equivalent to free electrodynamics of longitudinal and scalar photons, and
in this respect the state $|\Psi>$ conveys the information correspondent
to the classical case.  What is physically important is the eigenvalue of
the constant operator $M_0$, Eq.\ (53). The vacuum has a quantum number:
the eigenvalue of the mass operator. Thus the theory is reduced
essentially to quantum mechanics, while the rest is coordinate
reparametrization. One may conjecture that this mechanism is at the basis
of the dimensional reduction for all the quantum field models for which
classically the Birkhoff theorem holds. For the general model (1) the
problem is the existence and identification of the canonical
transformation, analogous to (12), that leads to free fields. 
\beginappendix{Linearization of the constraints}
The Hamiltonian and supermomentum constraints are quadratic in the
canonical coordinates and momenta. We will show that we have further
positivity information in the present case, that allows to linearize the
constraints.  It follows that the quantization of this model is not
equivalent to the quantization of the bosonic string. Let us see this in
detail. The constraints ${\cal H}_{\pm}=0$ can be cast in the form
$$\eqalign{&(\pi_0-\lambda A_0')^2=(\pi^1+\lambda A_1')^2\,,\cr
&(\pi_0+\lambda A_0')^2=(\pi^1-\lambda A_1')^2\,.\cr}\eqno(a.1)
$$
Classically Eqs.\ ($a.1$) can be satisfied if one of the following cases 
holds:
\item{$i$} $\pi^0-\pi^1=0$ and $A_0'+A_1'=0$;
\item{$ii$} ${\cal C}^{\alpha}=0$;
\item{$iii$} ${\cal D}^{\alpha}=0$; 
\item{$iv$} $\pi^0+\pi^1=0$ and $A_0'-A_1'=0$.

Case $ii$ coincides with the choice (26). For the bosonic string all four
cases are possible. In the present model the fields must satisfy the
relations (12) that select $ii$ as the only option.  Consider for example
the case $iii$ and use (12).  This is clearly inconsistent because it
implies $e^{f/2}\cosh\Sigma=0$.  Analogously, cases $i$ and $iv$ are not
consistent with the canonical transformation (12). As a consequence, the
linearized constraints ${\cal C}^{\alpha}=0$ of $ii$ are completely
equivalent to the constraints ($a.1$). 

One can ask why the linearization of the constraints is so ``asymmetric''
and the case 2) is the only consistent choice. (Why not $iii$?) The answer
to this question can be found in the canonical transformation (12). The
connection between the two-dimensional dilaton-gravity and the bosonic
string was derived on the assumption that $\rho$ is positive
($\rho=e^{f}$). This leads to the constraints ($5a,b$) and to Eqs.\
($14a,b$) via the transformation (12). Let us suppose for a moment that
$\rho=-e^{f}$. Since the Ricci scalar is an odd function of the metric,
the first three terms of $\cal H$ and $\cal P$ have opposite signs with
respect to ($5a,b$). (Alternatively, an overall opposite sign and
$\lambda\to -\lambda$.) In this case the canonical transformation leading
to the bosonic string reads
$$\eqalign{&A_0=-{2\over\lambda}e^{-f/2}\left(\pf\cosh{\Sigma}-
\vf'\sinh{\Sigma}\right)\,,~~
\pz=\lambda e^{f/2}\cosh{\Sigma}+\lambda A_1'\,,\cr
&A_1=-{2\over\lambda}e^{-f/2}\left(\pf\sinh{\Sigma}-
\vf'\cosh{\Sigma}\right)\,,~~
\pu=-\lambda e^{f/2}\sinh{\Sigma}-\lambda A_0'\,.\cr}\eqno(a.3)$$
Repeating the linearization of the constraints illustrated above it is
easy to see that the ``consistent linearized constraints'' are now
represented by the choice $iii$. In conclusion we can say that the choice
of the linearization corresponds to the sign of the two-dimensional
metric.

Let us go back to $\rho>0$ and conclude this discussion with an important
remark about the signs of ${\cal X}'_{\pm}$. From the canonical
transformation (12) it is straightforward to see that
$$\eqalign{&{\cal D}^0=-\lambda e^{f/2}\cosh\Sigma\,,\cr
&{\cal D}^1=e^{f/2}\sinh\Sigma\,.\cr}\eqno(a.4)$$
As a consequence we have ($\lambda$ is positive)
$$\eqalign{-2{\cal X}_-'&\equiv
{1 \over \sqrt{\lambda}}\,\bigl({\cal D}^0+{\cal D}^1\bigr)
=-\sqrt{\lambda}\,\,e^{f/2}e^{-\Sigma}<0\,,\cr
2{\cal X}_+'&\equiv{1 \over \sqrt{\lambda}}\bigl({\cal D}^0-{\cal
D}^1\bigr)
=-\sqrt{\lambda}\,e^{f/2}e^{\Sigma}<0\,.\cr}
\eqno(a.5)$$
The definite signs of both functions ${\cal X}_+'$, ${\cal X}_-'$ allows
them to be embedded into the Lagrange multipliers.
\beginack
This investigation was supported in part by the grants RFBR97-01-01041
and INTAS 93-127-ext. One of us (M.C.) is supported by a Human Capital and
Mobility grant of the European Union, contract number ERBFMRX-CT96-0012.
\beginref
\item{[1]} \Strom.
\item{[2]} \Fil.
\item{[3]} \CGHS.
\item{[4]} \CJA; ~\CJB; ~\CJC.
\item{[5]} \BJL; ~\CJZ.
\item{[6]} \KRV.
\item{[7]} \BOL.
\item{[8]} \LGK.
\item{[9]} \AER.
\item{[10]} \Strobl.
\item{[11]} \BH.
\item{[12]} \Nav.
\item{[13]} \Jacksm.
\item{[14]} \KK.
\item{[15]} \Land.
\item{[16]} \Kumb.
\item{[17]} \Jauch.
\vfill
\eject
\bye